\documentclass[twocolumn,floatfix,superscriptaddress,longbibliography]{revtex4-1}

\usepackage{amssymb,amsmath,bm}
\usepackage{graphicx}
\usepackage{epstopdf}
\usepackage{color}
\usepackage{appendix}
\usepackage[T1]{fontenc}
\usepackage[colorlinks=true,citecolor=blue,linkcolor=magenta]{hyperref}

\newcommand{\bbA}{\mathbb{A}}
\newcommand{\bbB}{\mathbb{B}}
\newcommand{\bbI}{\mathbb{I}}
\newcommand{\bbF}{\mathbb{F}}
\newcommand{\bbM}{\mathbb{M}}
\newcommand{\bbV}{\mathbb{V}}

\newcommand{\bbX}{\mathbb{X}}
\newcommand{\bbY}{\mathbb{Y}}
\newcommand{\bbZ}{\mathbb{Z}}

\newcommand{\mA}{\mathcal{A}}
\newcommand{\mB}{\mathcal{B}}
\newcommand{\mC}{\mathcal{C}}

\newcommand{\rmh}{\mathrm{h}}
\newcommand{\rmv}{\mathrm{v}}
\newcommand{\rmhv}{\mathrm{hv}}
\newcommand{\ket}[1]{\mbox{$| #1 \rangle$}}

\newcommand{\ingr}[2]{\begin{matrix}\includegraphics[height=#1 cm]{#2}\end{matrix}}

\newcommand\numberthis{\addtocounter{equation}{1}\tag{\theequation}}

\begin{document}

\title{Determining topological order from infinite projected entangled pair states}

\author{Anna Francuz}
\affiliation{Institute of Physics, Jagiellonian University, 
             {\L}ojasiewicza 11, PL-30348 Krak\'ow, Poland}

\author{Jacek Dziarmaga}
\affiliation{Institute of Physics, Jagiellonian University, 
             {\L}ojasiewicza 11, PL-30348 Krak\'ow, Poland}
             
\author{Guifre Vidal}
\affiliation{Perimeter Institute for Theoretical Physics, Waterloo, Ontario, N2L 2Y5 Canada}           
\affiliation{X, The Moonshot Factory, Mountain View, CA 94043}

\author{Lukasz Cincio}
\email[corresponding author: ]{lcincio@lanl.gov}
\affiliation{Theory Division, Los Alamos National Laboratory, Los Alamos, NM 87545}

\date{\today}

\begin{abstract}
We present a method of extracting information about the topological order from the ground state of a strongly correlated two-dimensional system computed with the infinite projected entangled pair state (iPEPS). For topologically ordered systems, the iPEPS wrapped on a torus becomes a superposition of degenerate, locally indistinguishable ground states. Projectors in the form of infinite matrix product operators (iMPO) onto states with well-defined anyon flux are used to compute topological $S$ and $T$ matrices (encoding mutual- and self-statistics of emergent anyons). The algorithm is shown to be robust against a perturbation driving string-net toric code across a phase transition to a ferromagnetic phase. Our approach provides accurate results near quantum phase transition, where the correlation length is prohibitively large for other numerical methods. Moreover, we used numerically optimized iPEPS describing the ground state of the Kitaev honeycomb model in the toric code phase and obtained topological data in excellent agreement with theoretical prediction. 
\end{abstract}

\maketitle

Topologically ordered phases \cite{wen1990topological} have in recent years attracted significant attention, mostly due to the fact that they support anyonic excitations --- exotic quasiparticles that obey fractional statistics. They are of interest not only from a fundamental perspective but also because of the possibility of realizing fault-tolerant quantum computation \cite{kitaev2003fault-tolerant} based on the braiding of non-Abelian anyons. An important challenge is to identify microscopic lattice Hamiltonians that can realize such exotic phases of matter. Apart from a number of exactly solvable models \cite{kitaev2003fault-tolerant, kitaev2006anyons, levin2005string}, verifying whether a given microscopic Hamiltonian realizes a topologically ordered phase and accessing its properties has traditionally been regarded as an extremely hard task.

A leading computational approach is to use Density Matrix Renormalization Group (DMRG) \cite{white1992density, white1993density} on a long cylinder \cite{yan2011spinliquid, jiang2012identifying, gong2013phase, zhu2013weak, gong2014emergent, zhu2014quantum, gong15global, hu2015topological, zhu15emergence, zhu2015spin, zaletel2016space, zeng2017nature, vaezi2017numerical, zhu2018robust}. In the limit of infinitely long cylinders, DMRG naturally produces ground states with well-defined anyonic flux, from which one can obtain full characterization of a topological order, via so-called topological $S$ and $T$ matrices \cite{cincio2013characterizing}. Since the proposal of Ref.~[\onlinecite{cincio2013characterizing}], the study of topological order by computing the ground states of an infinite cylinder with DMRG has become a common practice \cite{he2014chiral, zhu2014topological, zhu2014chiral, bauer2014chiral, zhu2015fractional, grushin15characterization, he2015kagome, he15distinct, he15bosonic,  geraedts15competing, mong2015fibonacci, he17realizing, stoudenmire2015assembling, he17signatures, Saadatmand20016symmetry, hickey2016haldane, zaletel2017measuring, zeng2018tunning}.

The cost of a DMRG simulation grows exponentially with the width of cylinder, effectively restricting this approach to thin cylinders. Instead, (infinite) Projected Entangled Pair States (iPEPS) allow for much larger systems \cite{verstraete2004renormalization, murg2007variational, verstraete2008matrix}. However, (variationally optimized) iPEPS naturally describe ground states with a superposition of anyonic fluxes. Here we show, starting with one such PEPS, how to produce a PEPS-like tensor network for each ground state with well-defined flux. Such tensor networks are suitable for extracting topological $S$ and $T$ matrices by computing overlaps between ground states. 

Our approach does not assume a clean realization of certain symmetries on the bond indices, in contrast to \cite{burak2014characterizing, bultinck2017anyons, iqbal2018study, fernandez2016constructing}. It also has much lower cost than methods based on the Tensor Renormalization Group \cite{he2014modular}. 

\begin{figure}[t!]
\includegraphics[width=\columnwidth]{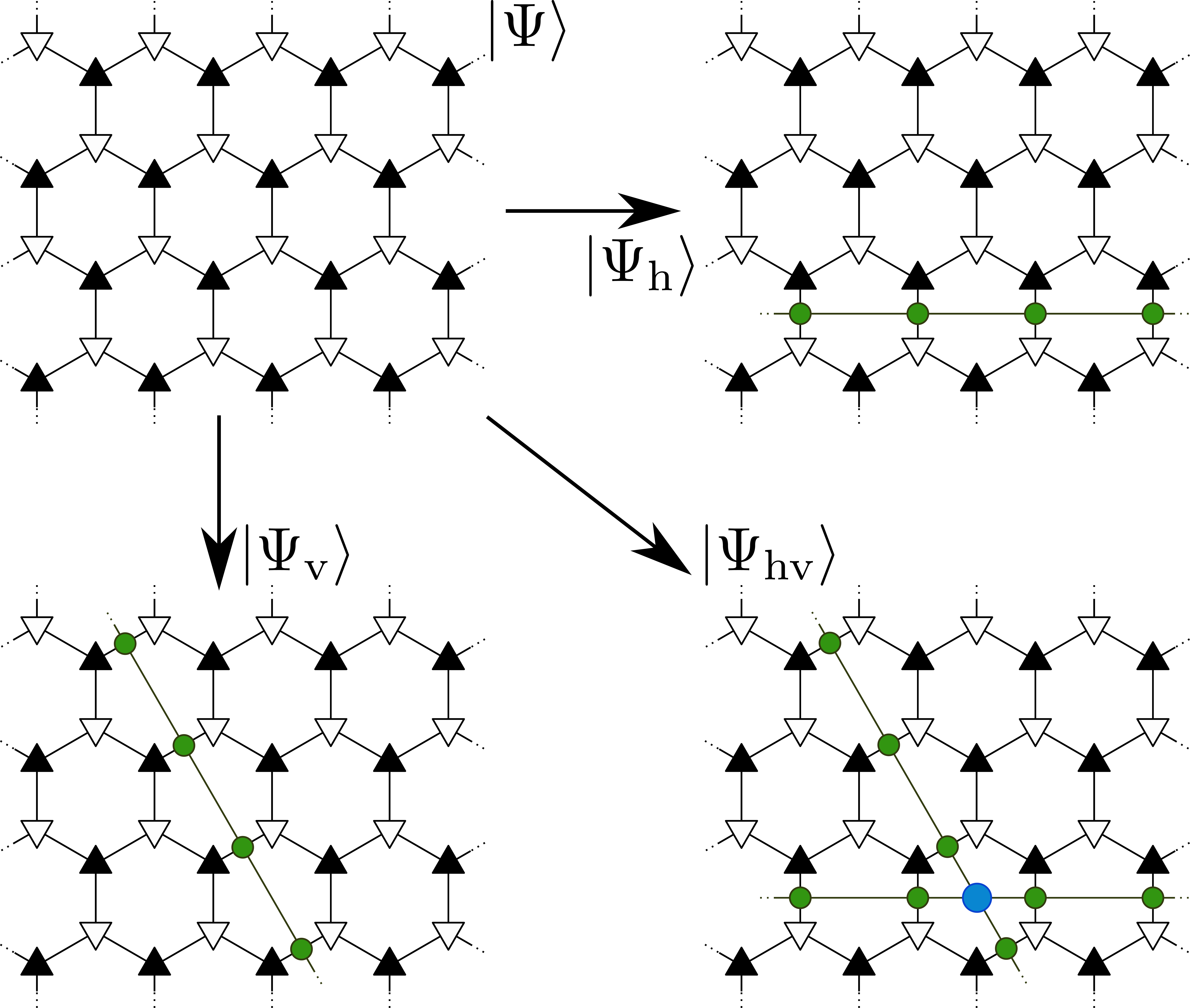}
\caption{A set of states $\ket{\Psi_\rmv}$, $\ket{\Psi_\rmh}$, $\ket{\Psi_\rmhv}$ is constructed from a single PEPS $\ket{\Psi}$ by inserting various MPOs in its bond indices. For a topologically ordered phase (toric code on a honeycomb lattice in this example), a proper combination of four states $\ket{\Psi}$, $\ket{\Psi_\rmv}$, $\ket{\Psi_\rmh}$ and $\ket{\Psi_\rmhv}$ is used to construct a basis of states with well-defined anyonic flux in a given direction. Physical indices are not drawn for simplicity. See text for details. }
\label{fig:summary}
\end{figure}

In this paper we employ variational method to minimize the energy of the iPEPS \cite{varCorboz}. The optimized state is then wrapped on a torus and the boundary conditions (with respect to the symmetry acting on the bond indices of PEPS) are suitably modified to recover all anyonic sectors. Figure~\ref{fig:summary} presents an overview of our approach. Computations are performed in the limit of an infinitely large torus allowing for accurate description of a topologically ordered phase even for models displaying a large correlation length. For clarity, we specialize the construction to PEPS describing the toric code realized by a string-net model on a honeycomb lattice \cite{levin2005string}. The method can be applied to other Abelian anyon models, as discussed below, and extended to non-Abelian ones \cite{AF_prep}.

In the toric code, the entanglement spectrum along the topologically nontrivial cut of a torus is supported on a vector space, which is a direct sum of four sectors, corresponding to the identity $\bbI$, bosonic $e$ and $m$ and fermionic~$\epsilon$ fluxes:
\begin{equation} \label{eq:TCspaces}
    \bbV^{\mathrm{TC}} = 
    \bbV^\bbI \oplus \bbV^e \oplus \bbV^m \oplus \bbV^\epsilon \ .
\end{equation}

We proceed by constructing projectors on ground states with definite anyon flux. The projectors are optimized and represented by matrix product operators (MPO). When inserted into PEPS and wrapped on the torus, the optimal MPO projects onto the desired ground state. Topological $S$ and $T$ matrices are extracted \cite{zhang2012quasiparticle, zhang2015general} by calculating overlaps between states with well-defined flux on tori related by modular transformations.

\begin{figure}[t!]
\includegraphics[width=\columnwidth]{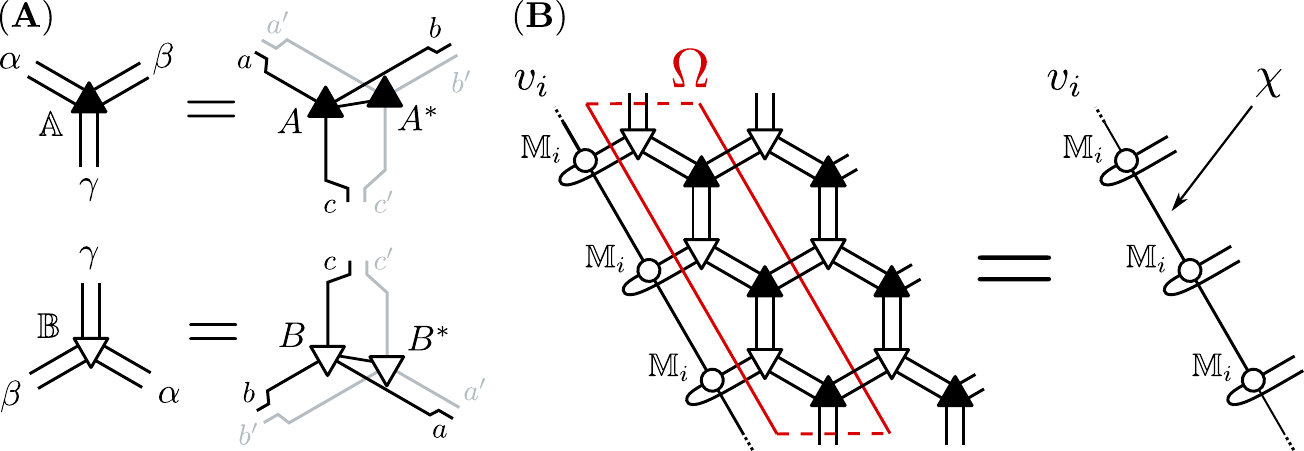}
\caption{(A) Graphical representation of double tensors $\bbA$ and $\bbB$.
(B) Left eigenvector $v_i$ of vertical TM $\Omega$ takes an MPO form. Vector $v_i$ is constructed with a single tensor $\bbM_i$ with bond dimension $\chi$, for $i=1,2$ and is obtained using a boundary MPS method described in detail in Appendix~\ref{app:tme}.
}
\label{fig:TM}
\end{figure}

{\it Transfer matrices and their eigenvectors. ---} PEPS for a toric code on a honeycomb lattice may be characterized by two tensors $A$ and $B$ with elements $A^i_{abc}$ and $B^i_{abc}$ respectively. Here, $i$ is a physical index and $a,b,c$ are bond indices. Let $\bbA$ and $\bbB$ denote double tensors $\bbA = \sum_i A^i \otimes (A^i)^\ast$ and $\bbB = \sum_i B^i \otimes (B^i)^\ast$ with double bond indices $\alpha = (a,a')$, etc., see Fig.~\ref{fig:TM}(A). PEPS transfer matrix (TM) $\Omega$ is defined by a line of tensors $\bbA$ and $\bbB$ contracted via some of their indices, as shown in Fig.~\ref{fig:TM}(B).

For a toric code PEPS we observe that $\Omega$ contains a~direct sum of $n=2$ topological sectors. Thus, the reduced density matrix on the virtual indices (which is directly related to the physical reduced density matrix \cite{cirac2011entanglement}) at a~topologically nontrivial cut is a~direct sum of two contributions
\begin{equation} \label{eq:vcut}
    \bbV_{\textrm{cut}} = \bbV^\bbI \oplus \bbV^e ~ \Rightarrow ~
    \rho_{\textrm{cut}} = \rho^\bbI \oplus \rho^e
\end{equation}
(recall that the ground state degeneracy of a toric code on a torus is $n^2 = 4$). The use of a pure MPS \cite{note_pure} as an ansatz for the dominant eigenvectors $v_1$, $v_2$ of $\Omega$ selects a~specific linear combination of sectors. We note that only the method based on boundary MPS (presented here) is capable of breaking the degeneracy of the dominant eigenvectors into minimally entangled states. Methods based on corner transfer matrix treat vertical and horizontal directions on the same footing and therefore will not select a minimally entangled state in a given direction. Numerically, eigenvectors $v_i$ may be obtained using a power method or by more advanced approaches such as the VUMPS algorithm \cite{tangent_review}, see Appendix~\ref{app:tme} for details. In the diagonal basis, they take the following form
\begin{equation} \label{eq:v_rho}
    v_1 = \rho^\bbI \oplus \rho^e \ , \quad v_2 = \rho^\bbI \oplus - \rho^e \ ,
\end{equation}
where we regard vector $v_i$ as an operator represented by an MPO constructed with a single tensor $\bbM_i$ as shown in Fig.~\ref{fig:TM}(B). Here, $\rho^\bbI$ and $\rho^e$ are boundary density matrices in identity and bosonic sectors, respectively. For clarity, we omitted the fact that vectors $v_i$ may contain a zero component, that is $v_1 = \rho^\bbI \oplus \rho^e \oplus 0$ and similarly for $v_2$. This leads to numerical instabilities and other complications that we discuss in detail in Appendix~\ref{app:tme}.

Matrix product description of $v_1$ and $v_2$ allows us to find an operator $Z_\rmv$ in the form of an MPO that maps $v_1$ into $v_2$ and back by demanding that
\begin{equation} \label{eq:Z}
    v_1 Z_\rmv = v_2 \ , \quad Z_\rmv v_2 = v_1 \ .
\end{equation}
In the diagonal basis of Eq.~\eqref{eq:v_rho}, $Z_\rmv = \bbI \oplus -\bbI$. We stress that we are able to obtain the generator of the global $\bbZ_2$ ``spin-flip'' symmetry that acts on the bond indices of PEPS, even though the symmetry is not realized on-site. In other words, PEPS tensors $A$ and $B$ do not have to be symmetric, as required in \cite{burak2014characterizing, bultinck2017anyons, iqbal2018study, fernandez2016constructing}, for our construction to work.

Similarly, we define horizontal TM $\Omega_\mathrm{h}$ and obtain its $n=2$ degenerate leading eigenvectors $h_1$ and $h_2$. Again, we are able to find an operator $Z_\rmh$ such that
\begin{equation} \label{eq:Zh}
    h_1 Z_\rmh = h_2 \ , \quad Z_\rmh h_2 = h_1 \ .
\end{equation}

\begin{figure}[t!]
\includegraphics[width=\columnwidth]{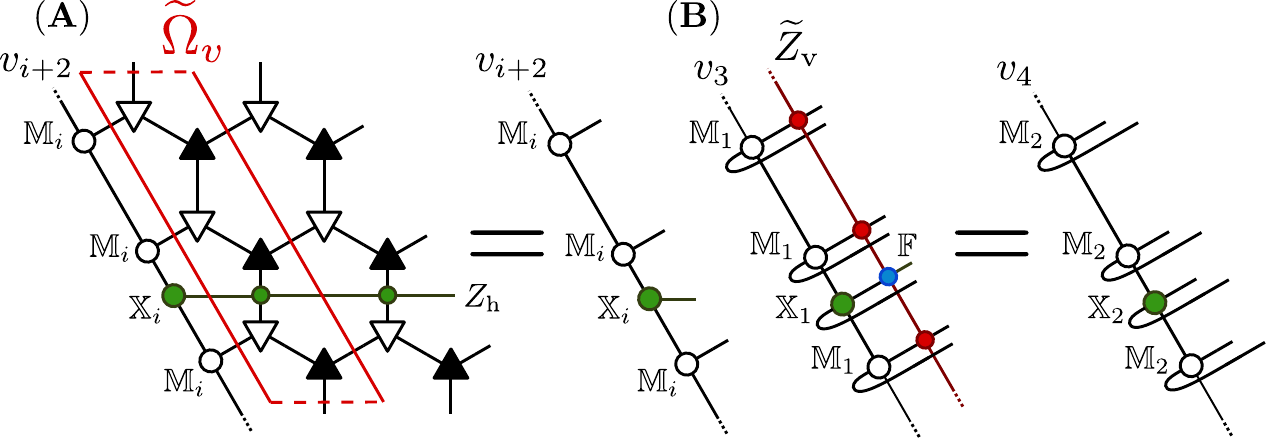}
\caption{(A)~Two eigenvectors $v_3$ and $v_4$ are obtained as pure MPOs from $v_1$ and $v_2$ by introducing additional tensors $\bbX_1$ and $\bbX_2$, which are obtained variationally, $i=1,2$. Tensors $\bbX_i$ are chosen such that $v_3$ and $v_4$ are leading eigenvectors of ``impurity'' TM $\widetilde{\Omega}_\rmv$. Double lines are dropped to improve clarity. (B) Graphical illustration of one of the conditions for $\widetilde{Z}_\rmv$ in Eq.~(\ref{eq:Ztilde}).}
\label{fig:impurityTM}
\end{figure}

Finally, we build vertical ``impurity'' TM $\widetilde{\Omega}_\rmv$ by inserting the $Z_\rmh$ operator on a horizontal cut of PEPS, as shown in Fig.~\ref{fig:impurityTM}(A). $Z_\rmh$ implements anti-periodic boundary conditions with respect to $\bbZ_2$ ``spin-flip'' symmetry acting in the PEPS bond indices. Note that, even if the $\bbZ_2$ symmetry is not realized on site, we still know that $Z_\rmh$ changes the boundary conditions from periodic to anti-periodic. Thus, inserting $Z_\rmh$ allows us to access two remaining sectors
\begin{equation} \label{eq:vcut_m_eps}
    \widetilde{\bbV}_{\textrm{cut}} = \bbV^m \oplus \bbV^\epsilon ~ \Rightarrow ~
    \widetilde{\rho}_{\textrm{cut}} = \rho^m \oplus \rho^\epsilon \ .
\end{equation}

As expected, we find $n=2$ leading eigenvectors of $\widetilde{\Omega}_\rmv$ that in some basis take the form
\begin{equation} \label{eq:v_rho_3_4}
    v_3 = \rho^m \oplus   \rho^\epsilon \ , \quad 
    v_4 = \rho^m \oplus - \rho^\epsilon \ .
\end{equation}

\begin{figure}[t!]
\includegraphics[width=0.8\columnwidth]{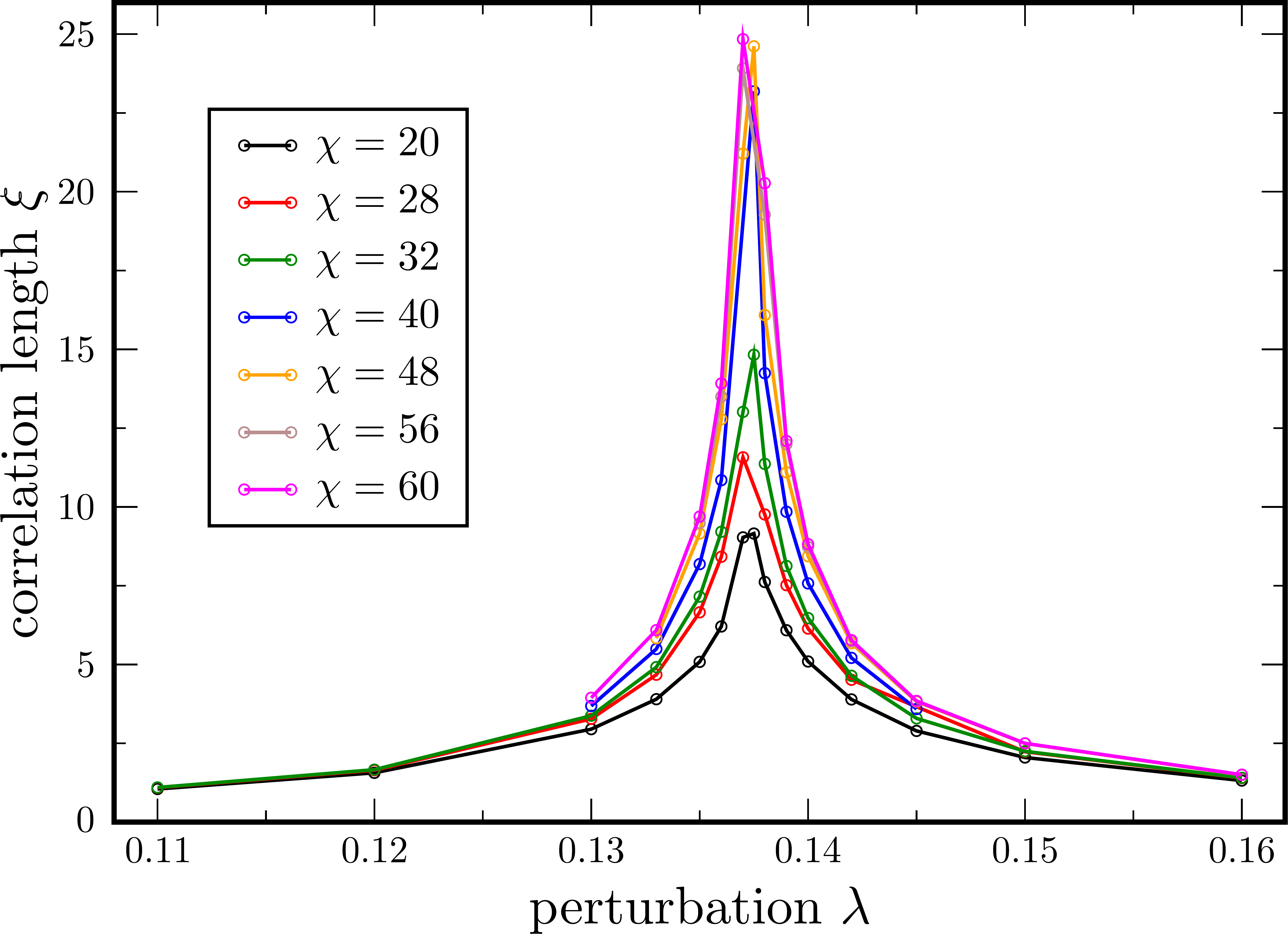}
\caption{
Correlation length $\xi$ as a function of perturbation strength $\lambda$ and a bond dimension $\chi$ of the TM eigenvectors $v_i$. Increasing $\chi$ reveals a quantum phase transition at $\lambda=0.137...0.138$. It separates toric code and ferromagnetic phases.
}
\label{corr_length}
\end{figure}

Eigenvectors $v_3$ and $v_4$ are obtained as pure MPOs~\cite{note_pure} from $v_1$ and $v_2$ by allowing for additional tensors $\bbX_i$, as depicted in Fig.~\ref{fig:impurityTM}(A). Note that tensors $\bbX_i$ are obtained variationally. In the limit of vanishing correlation length $\xi$ in the toric code PEPS studied here, the above ansatz for $v_3$ and $v_4$ becomes exact. In other models, bond dimension $\chi$ of all $v_i$ is increased to account for potentially large $\xi$. Our ansatz is validated by the results presented below. There, the correlation length $\xi \approx 25$ does not significantly impact the quality of the final result, see Fig.~\ref{corr_length} and the discussion below it. 

$\bbZ_2$ symmetry acting on the anti-periodic sectors is realized by an operator $\widetilde{Z}_\rmv$ satisfying
\begin{equation} \label{eq:Ztilde}
    v_3 \widetilde{Z}_\rmv = v_4 \ , \quad \widetilde{Z}_\rmv v_4 = v_3 \ .
\end{equation}
The construction of $\widetilde{Z}_\rmv$ mirrors that of $v_3$ and $v_4$. $\widetilde{Z}_\rmv$ is obtained from $Z_\rmv$ by allowing for additional variational tensor $\bbF$. Figure~\ref{fig:impurityTM}(B) shows one condition from Eq.~\eqref{eq:Ztilde} that is used to compute $\bbF$. $\bbF$ is one of the generators of $C^*$-algebra, from which central idempotents can be found~\cite{bultinck2017anyons}.

Appendix~\ref{app:tme} details some numerical issues associated with finding vectors $v_i$, $i=1,\ldots,4$ as well as solving Eqs.~(\ref{eq:Z}) and~(\ref{eq:Ztilde}).

{\it Projectors onto definite anyon fluxes. ---} Symmetry group generators $Z_\rmv$ and $\widetilde{Z}_\rmv$ can be used to construct ground states with well-defined flux in the horizontal direction. Recall that $Z_\rmv$ realizes $\bbZ_2$ symmetry in the periodic sector $\bbV^\bbI \oplus \bbV^e$. Operators $P^\pm = (\bbI \pm Z_\rmv)/2$ are thus projectors on definite anyonic sectors and states 
\begin{equation}
\ket{\Psi^\bbI} \sim  \ket{\Psi} + \ket{\Psi_\rmv} \ , \quad
\ket{\Psi^e} \sim \ket{\Psi} - \ket{\Psi_\rmv}
\end{equation}
have well-defined identity and electric flux in the horizontal direction, respectively. Note that projectors $P^\pm$ do not act on the physical Hilbert space. Instead, they are defined on the bond indices of PEPS. The above construction is summarized in Fig.~\ref{fig:summary}. Here, $\ket{\Psi}$ denotes the initial PEPS state and $\ket{\Psi_\rmv}$ is the state obtained by inserting $Z_\rmv$ into bond indices of PEPS that defines $\ket{\Psi}$. We remark that projectors $P^\pm$ play the same role as projector MPO's in the construction of MPO-injective PEPS \cite{bultinck2017anyons}. 

Similarly, $\widetilde{Z}_\rmv$ generates the $\bbZ_2$ symmetry group in the anti-periodic sector $\bbV^m \oplus \bbV^\epsilon$. It defines projectors $\widetilde{P}^\pm = (\bbI \pm \widetilde{Z}_\rmv)/2$. States with well-defined magnetic $\ket{\Psi^m}$ and fermionic $\ket{\Psi^\epsilon}$ flux are obtained by first changing the boundary conditions on the bond indices with $Z_\rmh$ and then projecting onto the proper subspace. That is, 
\begin{equation}
\ket{\Psi^m} \sim \ket{\Psi_\rmh} + \ket{\Psi_\rmhv} \ , \quad
\ket{\Psi^\epsilon} \sim \ket{\Psi_\rmh} - \ket{\Psi_\rmhv} \ ,
\end{equation}
where $\ket{\Psi_\rmh}$ stands for $\ket{\Psi}$ with $Z_\rmh$ inserted and $\ket{\Psi_\rmhv}$ denotes $\ket{\Psi_\rmh}$ that has $\widetilde{Z}_\rmv$ embedded in together with the tensor~$\bbF$. Figure~\ref{fig:summary} summarizes the construction of $\ket{\Psi_\rmh}$ and $\ket{\Psi_\rmhv}$.

{\it Topological $S$ and $T$ matrices. ---} States $\ket{\Psi^i}$ with well-defined flux $i = \bbI, e, m, \epsilon$ are used to calculate topological $S$ and $T$ matrices. The $T$ matrix is diagonal and stands for self-statistics, while the $S$ matrix encodes mutual statistics. Together they form a representation of a modular group $SL(2,\mathbb{Z})$, by which they are related to the modular transformations of a torus generated by $\mathfrak{s}$ and $\mathfrak{t}$ transformations \citep{wen2015theory}. It follows that overlaps between $\ket{\Psi^i}$ transformed by a combination of modular transformations $\mathfrak{s}$ and $\mathfrak{t}$ constitute entries of a corresponding combination of topological $S$ and $T$ matrices.

\begin{figure}[t!]
\includegraphics[width=0.7\columnwidth]{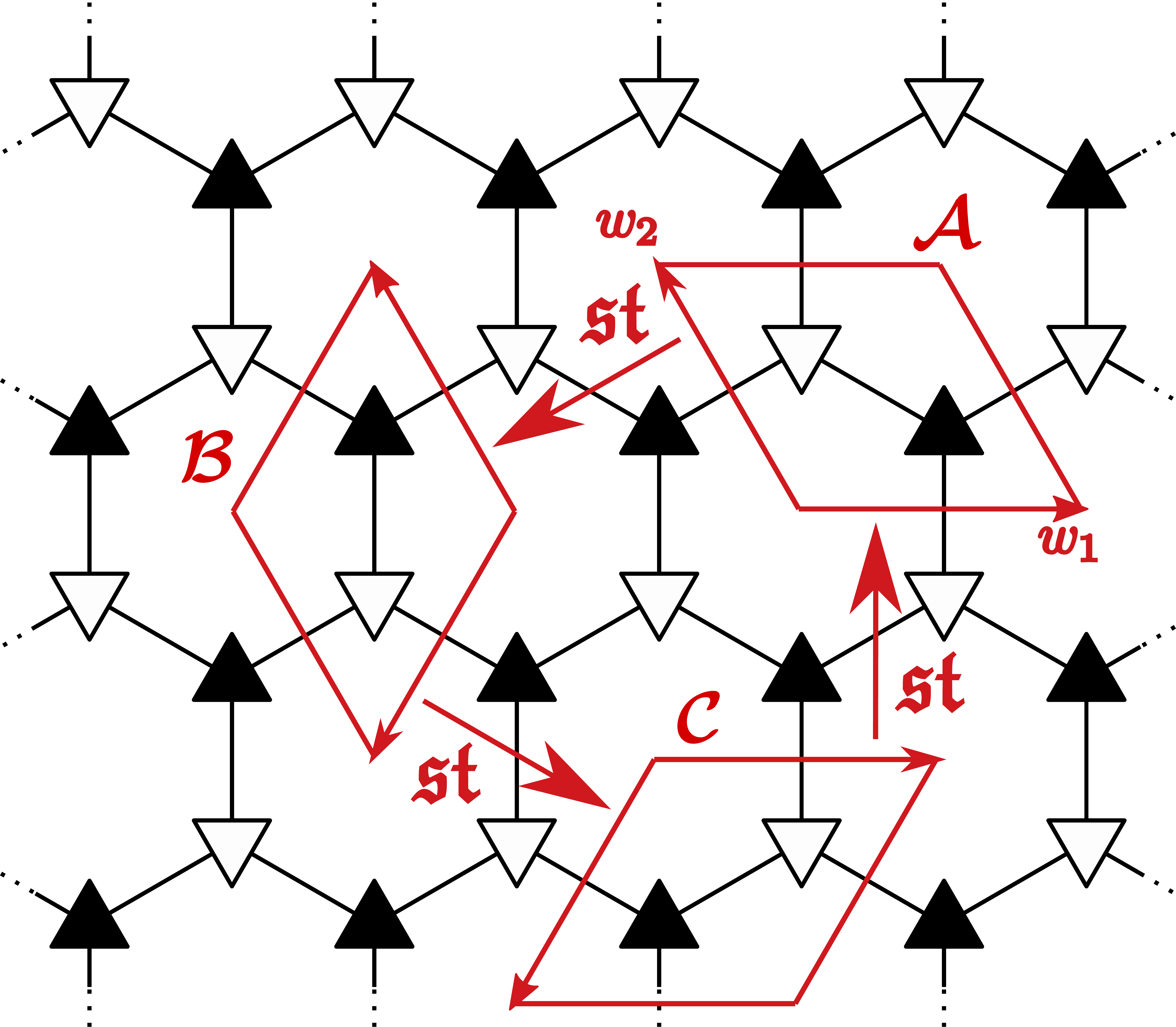}
\caption{Three tori $\mA$, $\mB$, and $\mC$ on a honeycomb lattice considered in our method. Torus $\mA$ is defined by a pair of vectors $(w_1,w_2)$. Each torus is obtained by $\mathfrak{st}$ modular transformation from another torus. Transformation $\mathfrak{st}$ corresponds to $120^\circ$ rotation, $(\mathfrak{st})^3~=~\bbI$. The described approach requires $120^\circ$ rotation symmetry of the lattice. Generalization to other symmetries is straightforward. Physical indices are not drawn for simplicity.}
\label{3tori}
\end{figure}

Throughout this paper, for concreteness, we work with the transformations on a lattice with  $120^\circ$ rotational symmetry. The construction is however general and applicable to lattices with other symmetries as well. We start by defining torus $\mA$ in Fig.~\ref{3tori} with unit vectors $w_1$, $w_2$ and corresponding transfer matrices: vertical $(w_1,N_vw_2)$ and horizontal $(N_hw_1,w_2)$, see Fig.~\ref{fig:TM}(B) for comparison. Similarly, we consider tori $\mB$ and $\mC$ together with their corresponding transfer matrices as shown in Fig.~\ref{3tori}.

Our method requires finding three complete sets of ground states 
\begin{equation} \label{eq:complete}
\left\{ \ket{\Psi^i_\mA} \right\}, \quad
\left\{ \ket{\Psi^i_\mB} \right\}, \quad
\left\{ \ket{\Psi^i_\mC} \right\}, \quad
i = \bbI, e, m, \epsilon
\end{equation}
with well-defined anyon fluxes corresponding to three different tori: $\mA$, $\mB$, $\mC$. Each torus is related to the previous one by a modular transformation $\mathfrak{st}$, which generates $120^\circ$ counterclockwise rotation, see Fig.~\ref{3tori}. Topological $S$ and $T$ matrices are extracted from all possible overlaps between states in (\ref{eq:complete}). This computation is presented in \cite{zhang2015general} and described in Appendix~\ref{app:STnoSym}. We stress that the presented method does not require any rotational invariance of the iPEPS tensors. 

{\it Toric code versus double semion and quantum double of $\mathbb{Z}_3$. ---} PEPS tensors that represent ground states of string-net models on a honeycomb lattice with zero correlation length can be found analytically \cite{buerschaper2009explicit, levin2005string-net}. As a~proof of principle, we numerically obtain topological $S$ and $T$ matrices for the toric code and the double semion model. Moreover, the described method gave exact $S$ and $T$ matrices for the quantum double of $\mathbb{Z}_3$ model defined on a square lattice \cite{schulz2012breakdown}. In this paper we restrict the description to the toric code phase realized in the (i) perturbed string-net model and (ii) Kitaev honeycomb model for which we analyze the iPEPS ground state obtained by numerical energy optimization.

{\it Perturbed string-net model. ---} In order to drive the iPEPS away from the fixed point with zero correlation length, we apply a perturbation $e^{-\lambda V}$ towards a ferromagnetic phase similarly as in \cite{per1,per2,per3} but with a two-site interaction
$V=-\sum_{\langle i,j \rangle}\sigma^x_i\sigma^x_j$, see Appendix~\ref{app:sn}.

Our method allows us to obtain accurate results even close to the critical point, in the regime of very long correlation lengths $\xi$, see Fig.~\ref{corr_length}. Indeed, for $\lambda=0.136$, where $\xi\approx 25$, we obtain $S = S_{\rm{tc}} + \epsilon_S$, $T = T_{\rm{tc}} + \epsilon_T$, where:
\begin{equation}
{\scriptsize S_{\rm{tc}} = \frac{1}{2}
\begin{pmatrix}
 1 & 1 & 1 & 1\\
  1 & 1 & -1 & -1\\
  1 & -1& 1& -1 \\
  1 & -1 & -1 & 1 \\
 \end{pmatrix},\qquad
 \scriptsize
  T_{\rm{tc}} = \begin{pmatrix}
  1 & 0 & 0 & 0 \\
  0 & 1 & 0 & 0 \\
  0 & 0 & 1 & 0 \\
  0 & 0 & 0 & -1 \\
 \end{pmatrix}} \ .
\end{equation}
The maximal element of $|\epsilon_S|$ and $|\epsilon_T|$ is of the order of $10^{-3}$ and $10^{-8}$, respectively.

In the ferromagnetic phase we find two eigenvectors of TM $\Omega$, $v_1 = \rho_1 \oplus 0$ and $v_2 = 0 \oplus \rho_2$. However, in contrast to the topologically ordered phase described by Eq.~(\ref{eq:Z}), there is no operator that maps $v_1$ to $v_2$. Numerically, this situation is detected by monitoring the distance (per lattice site) between $v_1 Z_\rmv$ and $v_2$. In the topologically trivial phase the distance converges to a finite value with growing bond dimension of $v_i$. 

\begin{figure}[t!]
\includegraphics[width=0.8\columnwidth]{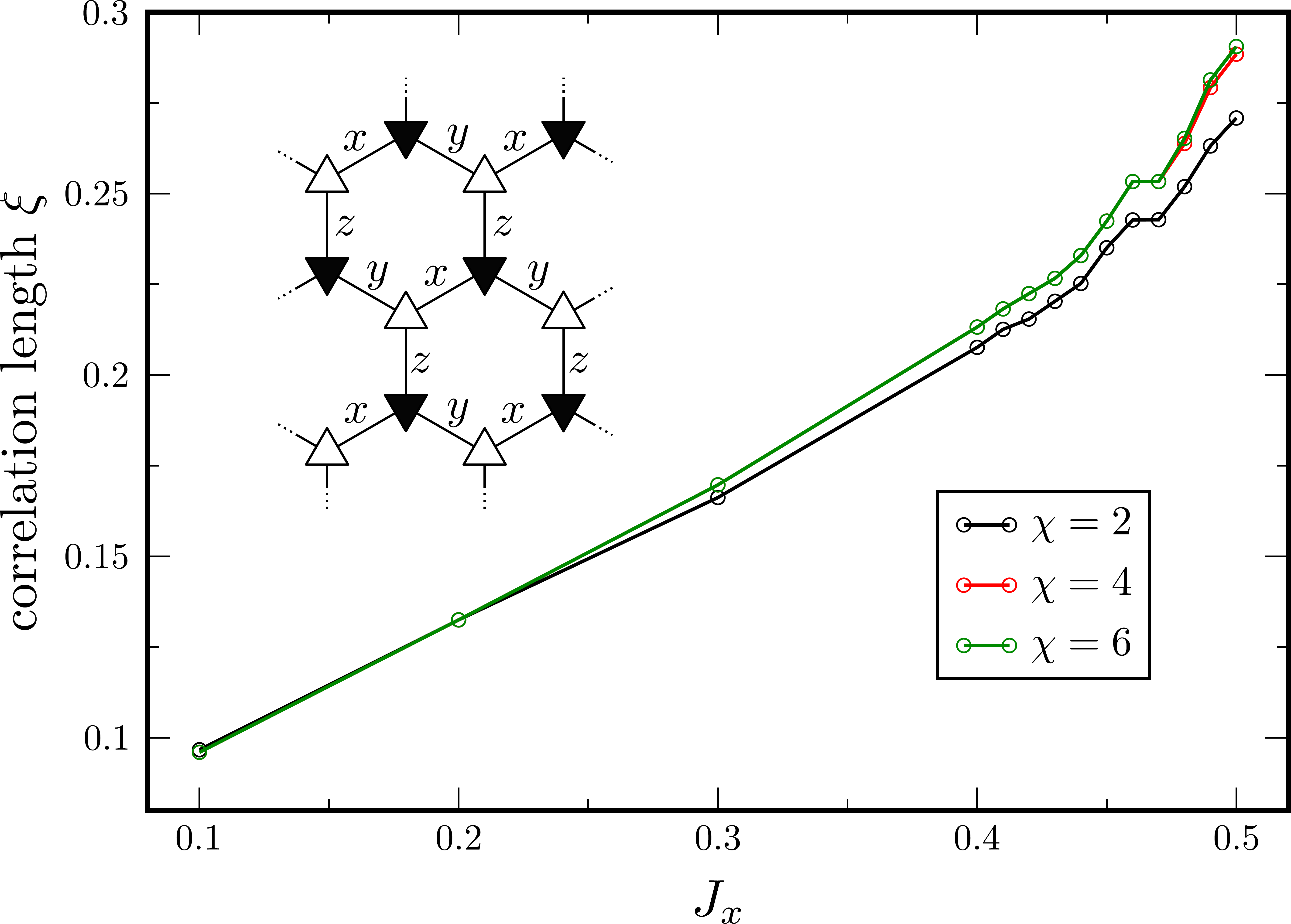}
\caption{
Correlation length $\xi$ as a function of $J_x = J_y$ in the Kitaev honeycomb model. Results for several values of bond dimension $\chi$ of TM eigenvectors $v_i$ are shown. Inset: graphical illustration of the Hamiltonian defined in Eq.~(\ref{eq:H_kit}) displaying three different types of coupling: $x$, $y$ and $z$. 
}
\label{fig:kitaev_CL}
\end{figure}

{\it Kitaev honeycomb model. ---} The model is defined by the following Hamiltonian
\begin{equation} \label{eq:H_kit}
    \mathcal{H} = - \sum_{\alpha=x,y,z} J_\alpha \sum_{\alpha~\rm{links}} \sigma^\alpha_i \sigma^\alpha_j
\end{equation}
on a honeycomb lattice. Here, $\sigma^\alpha_i$, $\alpha=x,y,z$ are Pauli matrices acting on site $i$. We set $J_z = 1$ and study the model along the line $J_x = J_y \in (0,0.5)$, see Fig.~\ref{fig:kitaev_CL}. The iPEPS ground state is obtained using variational optimization. We find that the bond dimension $\chi=4$ of boundary MPO's $v_i$ suffices to faithfully capture the entanglement properties of the phase.

We obtain correct topological $S$ and $T$ matrices within very small error. We are able to uniquely determine the anyon model for a range of parameters $J_x=J_y \in [0.2,0.48]$. Most notably for $J_x=J_y=0.44$, which is close to the critical point at $J_x=J_y=0.5$, we compute topological matrices $S = S_{\rm{tc}} + \epsilon_S$, $T = T_{\rm{tc}} + \epsilon_T$, where the maximal element of $|\epsilon_S|$ ($|\epsilon_T|$) is $1.3 \times 10^{-3}$ ($2.2 \times 10^{-3}$). The errors $|\epsilon_S|$, $|\epsilon_T|$ grow with increasing $J$, however stay below $4\%$ in the interval $J_x=J_y \in [0.2,0.48]$. This accuracy is sufficient to unambiguously determine the type of topological order.

{\it Conclusions. ---} We presented a method of identifying topological order from a microscopic lattice Hamiltonian that does not have explicit limitations on the size of the system. The method is based on extracting topological $S$ and $T$ matrices from a single iPEPS. Our techniques allow us to analyze systems with much bigger correlation length than the state-of-the-art 2D DMRG. Finally, we analyzed numerically optimized iPEPS describing the ground state of the Kitaev honeycomb model in the toric code phase. This computation shows that our approach does not require an artificially implemented realization of topological symmetries. Instead, it is applicable to generic, variationally obtained iPEPS.

{\it Acknowledgments. ---} AF would like to thank Jutho Haegeman, Frank Verstraete, Robijn Vanhove and Laurens Lootens for explaining their work \cite{bultinck2017anyons}. Numerical calculations were performed in MATLAB with the help of the \verb+ncon+ function \cite{NCON} for tensor contractions. This research was supported by the Polish Ministry of Science and Education under grant DI2015 021345 (AF) and by Narodowe Centrum Nauki (NCN) under grant 2016/23/B/ST3/00830 (AF, JD). LC was supported initially by the U.S. DOE through the J. Robert Oppenheimer fellowship and subsequently by the DOE, Office of Science, Basic Energy Sciences, Materials Sciences and Engineering Division, Condensed Matter Theory Program. GV is a CIFAR fellow in the Quantum Information Science Program. This research was supported in part by Perimeter Institute for Theoretical Physics. Research at Perimeter Institute is supported by the Government of Canada through the Department of Innovation, Science and Economic Development Canada and by the Province of Ontario through the Ministry of Research, Innovation and Science. X is formerly known as Google[x] and is part of the Alphabet family of companies, which includes Google, Verily, Waymo, and others (www.x.company).

%\bibliography{refs.bib}


\clearpage
\newpage

\appendix
\setcounter{page}{1}
\renewcommand\thefigure{\thesection.\arabic{figure}}
\setcounter{figure}{0}

\onecolumngrid
\begin{center}
\large{ Supplemental material for \\ ``Determining topological order from infinite projected entangled pair states''}
\end{center}

\twocolumngrid

\section{PEPS and topological order} \label{app:tme}

In this Appendix we give further details on numerical methods used to extract topological order from iPEPS that were omitted for clarity in the main text.

\subsection{PEPS transfer matrices and their eigenvectors}

We start by discussing some numerical issues that may be encountered in the analysis of PEPS transfer matrices and their eigenvectors.

We assume that leading eigenvectors of PEPS transfer matrices can be accurately approximated by MPOs. The approximation is controlled by a bond dimension $\chi$ of the eigenvector. For a topologically ordered state, the PEPS transfer matrix admits degenerate leading eigenvalue. Because of the topological nature of the degeneracy, it has to be exact even away from the zero-correlation fixed point. However, due to the assumption that the eigenvectors are constrained to be MPO, this degeneracy may be lifted. Indeed, this is what we observe numerically for PEPS with nonvanishing correlation length. The degeneracy is restored in the limit of large bond dimension of the MPO eigenvector.

The degeneracy of the leading eigenvalue of a PEPS transfer matrix is in general unknown. In practice, we can only bound it from below and proceed to the next steps of the approach with the assumption that all leading eigenvectors were obtained. This assumption is later verified by various consistency checks (unitarity of computed $S$ matrix, ability to find symmetry generators $Z$, etc.).

The above issue is analogous to the one observed in 2D iDMRG procedure of extracting topological order from microscopic lattice Hamiltonian \cite{cincio2013characterizing}. There, for (initially) unknown topological order, the number of degenerate ground states on an infinite cylinder is also unknown.

\begin{figure}[t!]
\includegraphics[width=0.9\columnwidth]{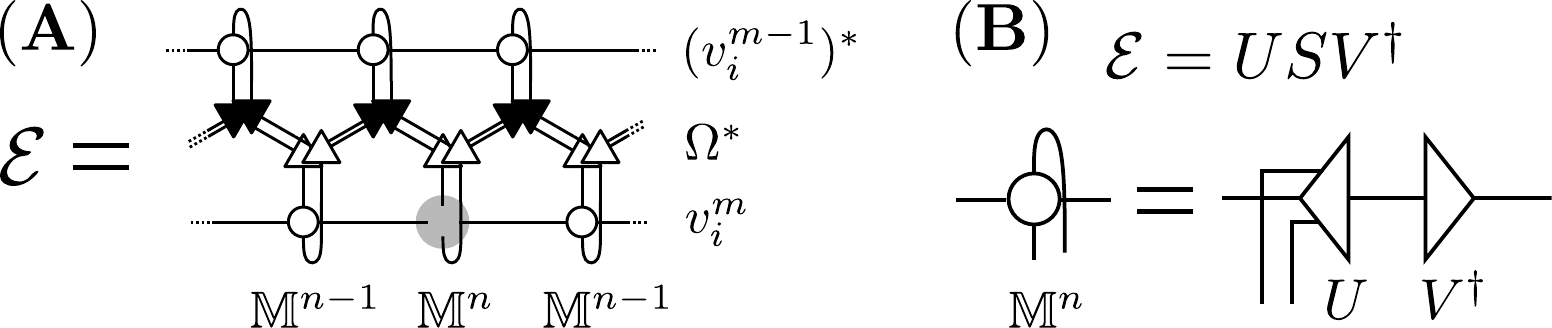}
\caption{Details on the algorithm used to find horizontal eigenvectors of PEPS transfer matrix. (A) The environment $\mathcal{E}$ of tensor $\mathbb{M}^n$ is computed from the overlap between MPO vectors $v_i^m$ and $\Omega^\ast v_i^{m-1}$. (B) The tensor $\mathbb{M}^n$ is updated by singular value decomposition of the environment $\mathcal{E}$.
}
\label{Algoritm_Eig}
\end{figure}

The eigenvectors to the leading eigenvalues of the PEPS transfer matrix are found with Lanczos power method. The resulting canonical MPS vector is a fixed point of a PEPS transfer matrix. In the Lanczos method, that fixed point is reached by iterative application of PEPS transfer matrix to the MPS vector. The above iteration is terminated when the vector $v_i$ in the $m$-th iteration is equal to the one from the previous iteration ($m-1$). Infinite MPS vectors (written in canonical form) are considered to be equal when their Schmidt coefficients are equal. In the $m$-th iteration, the new tensor $\mathbb{M}^n$ in the MPS vector $v_i^m$ is updated by maximizing the overlap with its environment $\mathcal{E}$, see Fig.~\ref{Algoritm_Eig}. The new tensor is given by singular decomposition of $\mathcal{E}$, which leads to an MPS in a the canonical form.
\begin{equation}
\mathcal{E} = U S V^\dagger, \quad \mathbb{M}^{n} = U V^\dagger.
\end{equation}
The environment $\mathcal{E}$ is formed from the overlap \mbox{$\langle \Omega\cdot v_i^{m-1}\vert v_i^m\rangle$}, with one tensor from the ket layer removed, as shown in the Fig.~\ref{Algoritm_Eig}. Once the $n$-th MPS tensor $\mathbb{M}^n$ is found, the environment is updated and the procedure is repeated for $\mathbb{M}^{n+1}$.

We noticed \cite{AF_prep} that the VUMPS algorithm \cite{tangent_review} converges to the same eigenvectors as our power method but typically requires fewer iterations.

\begin{figure}[t!]
\includegraphics[width=0.9\columnwidth]{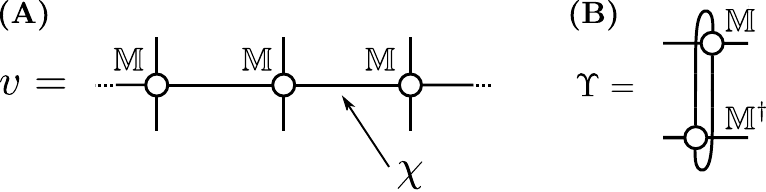}
\caption{(A) Eigenvector $v$ of a PEPS transfer matrix. It takes an MPO form with tensor $\bbM$. (B) Transfer matrix $\Upsilon$ built with tensor $\bbM$. The spectrum of $\Upsilon$ reveals whether the MPO eigenvector $v$ is pure and can be used to compute the correlation length in the PEPS state. See text for details.
}
\label{fig:MPOtm}
\end{figure}

Once the eigenvectors are computed, we analyze their transfer matrices $\Upsilon$. This construction is shown in Fig.~\ref{fig:MPOtm}. We say that a given MPO (or analogously, MPS) is pure, when the leading eigenvector of its corresponding transfer matrix is not degenerate. All eigenvectors of PEPS transfer matrices are required to be pure for our method to work. We note that this requirement is not always straightforward to meet. For a PEPS with small correlation length, the eigenvector MPO may carry contributions from several pure eigenvectors, if the computation is carried out with bond dimension that is too large. Correlation length shown in Fig.~\ref{corr_length} in the main text is computed by analyzing the second leading eigenvalue of the transfer matrix $\Upsilon$ shown in Fig.~\ref{fig:MPOtm}.

The above remarks are also relevant for vertical PEPS transfer matrix and the ones with impurity.

\subsection{Eigenvector decomposition}

As mentioned in the main text, the leading eigenvectors of PEPS transfer matrices admit decomposition involving boundary density matrices in given topological sectors. In most cases however, we expect that this decomposition will also involve a trivial, zero component that will be present due to redundancy in PEPS description of topologically ordered state. That is, we expect (and have numerically observed in the examples presented in the main text) that leading eigenvectors of a~vertical transfer matrix $\Omega$ in Fig.~\ref{fig:TM}(B) will take the form
\begin{equation}
    v_1 = \rho^\bbI \oplus \rho^e \oplus 0 \ , \quad v_2 = \rho^\bbI \oplus - \rho^e \oplus 0 \ ,
\end{equation}
where $0$ is the null operator in the complement of the supports of $\rho^\bbI$ and $\rho^e$, compare with Eq.~(\ref{eq:v_rho}) in the main text. Similarly, leading eigenvectors of impurity transfer matrix $\widetilde{\Omega}$ in Fig.~\ref{fig:impurityTM}(A) will be given by
\begin{equation}
    v_3 = \rho^m \oplus   \rho^\epsilon \oplus 0 \ , \quad 
    v_4 = \rho^m \oplus - \rho^\epsilon \oplus 0 \ ,
\end{equation}
see Eq.~(\ref{eq:v_rho_3_4}) for comparison.

\subsection{Projectors onto well-defined anyon fluxes}

\begin{figure}[t!]
\includegraphics[width=\columnwidth]{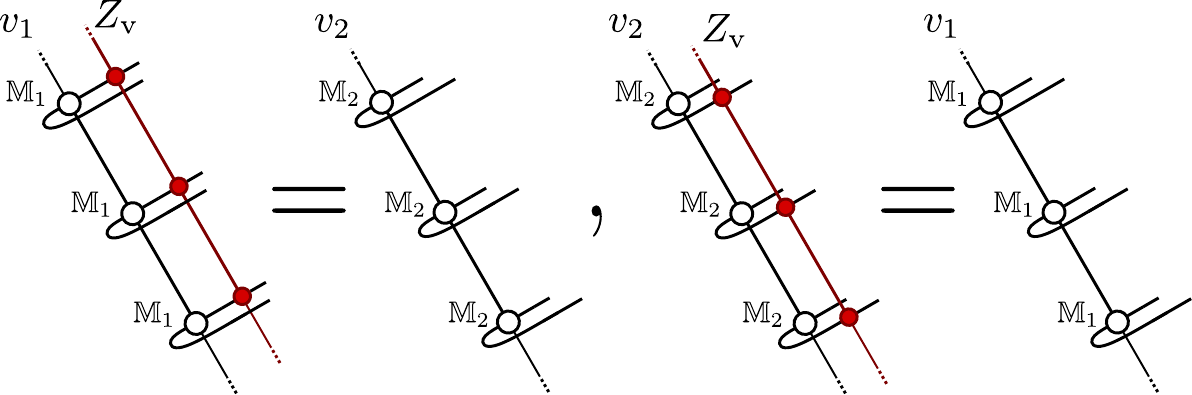}
\caption{Graphical illustration of two conditions for operator $Z_\rmv$ in Eq.~(\ref{app:Z}).
}
\label{fig:appZeq}
\end{figure}

The presence of the trivial component in eigenvectors $v_i$ impacts the process of finding the operator $Z_\rmv$ from
\begin{equation} \label{app:Z}
    v_1 Z_\rmv = v_2 \ , \quad Z_\rmv v_2 = v_1
\end{equation}
and thus affects finding the projectors \mbox{$P^{\pm} = (\bbI + Z_\rmv)/2$} onto states with well-defined anyon fluxes. Figure~\ref{fig:appZeq} shows graphical representation of both conditions in~(\ref{app:Z}). 

Since $v_i$ in Eqs.~(\ref{app:Z}) are not invertible, we attempt to find $Z_\rmv$ variationally. The corresponding optimization problem reads
\begin{equation} \label{app:opt}
    \min_z \left( || v_1 z - v_2 ||^2 + || z v_2 - v_1 ||^2  \right) \ ,
\end{equation}
where the optimization is performed over translation invariant MPOs. To solve that optimization problem, we employ techniques similar to the ones developed for iDMRG \cite{mcculloch08infinite}. Namely, the optimization starts by solving small system with open boundary conditions. The system is then slowly grown until the thermodynamic limit is reached. As in standard iDMRG, only few unit cells of tensors are explicitly stored in computer's memory and the system is grown by properly reevaluating boundary tensors. This method, when the growth of the system is done slowly enough, avoids numerical instabilities caused by the zero modes and solves the problem of degenerate solution space. 

In a standard iDMRG, a single tensor is optimized according to an energy minimization. In order to solve the optimization problem (\ref{app:opt}), we replace energy minimization part of the iDMRG in the following way. Let $z_0$ denote the tensor that undergoes the optimization. 
Then the expression in (\ref{app:opt}) can be written as
\begin{equation}
|| v_1 z - v_2 ||^2 + || z v_2 - v_1 ||^2 = 
z_0^\dag N z_0 - y^\dag z_0 - z_0^\dag y + c \ .  
\end{equation}
Here $z_0$ is vectorized, $N=N^\dag\geq0$ is an environment matrix, $y$ is a vector, and $c$ is a constant. This expression is minimized by $z_0$ that satisfies a linear equation:
\begin{equation} \label{eq:Mzy}
N z_0 = y \ .
\end{equation}
Equation (\ref{eq:Mzy}) yields a new tensor $z'_0 = N^{-1}y$, where the pseudoinverse $N^{-1}$ is obtained by singular value decomposition of matrix $N$ and carefully controlling the number of singular values which are inverted. 

In an alternative scheme, used occasionally, a new tensor $z^\mathrm{opt}$ that replaces $z$ in the whole MPO is given by a combination 
\begin{equation}
    z^\mathrm{opt} = \cos(\alpha) z_0 + \sin(\alpha) z'_0 \ ,
\end{equation}
where $\alpha$ minimizes the expression in (\ref{app:opt}). Optimal $\alpha$ is found using the procedure described in \cite{varCorboz}. 

\begin{figure}[t!]
\includegraphics[width=\columnwidth]{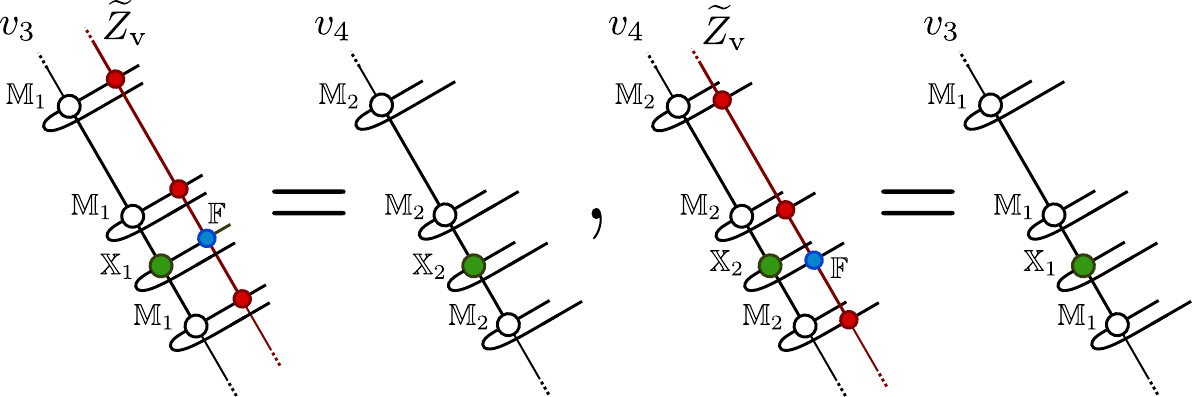}
\caption{Graphical illustration of two conditions for operator $\widetilde{Z}_\rmv$ in Eq.~(\ref{app:Ztilde}).
}
\label{fig:appZteq}
\end{figure}

Finally, let us comment on finding $\widetilde{Z}_\rmv$ from 
\begin{equation} \label{app:Ztilde}
    v_3 \widetilde{Z}_\rmv = v_4 \ , \quad \widetilde{Z}_\rmv v_4 = v_3 \ .
\end{equation}
Figure~\ref{fig:appZteq} shows graphical representation of both of the above equations. As noted in the main text, $\widetilde{Z}_\rmv$ can be obtained from $Z_\rmv$ by allowing for additional tensor $\bbF$. Because of that, the corresponding optimization procedure
\begin{equation}
    \min_z \left( || v_3 z - v_4 ||^2 + || z v_4 - v_3 ||^2  \right)
\end{equation}
is much simpler than (\ref{app:opt}) as the optimization is carried out over a single copy of a tensor (blue tensor in Fig.~\ref{fig:appZteq}). This is in contrast with optimization in (\ref{app:opt}), where the entire MPO (consisting of infinitely many copies of a given tensor) has to be found.

\setcounter{figure}{0} 
\section{General scheme of calculating topological $S$ and $T$ matrices without the presence of rotational symmetry} \label{app:STnoSym}

In this Appendix we discuss in detail the computation of topological $S$ and $T$ matrices from iPEPS. For concreteness, we describe the method using toric code example. The method itself is general and could be applied to other anyon models.

Our starting point is three sets of ground states with well-defined anyon flux on three tori $\mA$, $\mB$, $\mC$ given by PEPS with additional MPO operators, as discussed in the main text. The method begins with computing all possible overlaps between ground states:
\begin{equation} \label{eq:KLM}
K_{ij} = \langle \Psi^i_\mA | \Psi^j_\mB \rangle , \
L_{ij} = \langle \Psi^i_\mB | \Psi^j_\mC \rangle , \
M_{ij} = \langle \Psi^i_\mC | \Psi^j_\mA \rangle ,
\end{equation}
where $\ket{\Psi^i_\alpha}$ is the ground state with well-defined flux $i=\bbI, e, m, \epsilon$ on torus $\alpha = \mA,\mB,\mC$. Tori $\mA$, $\mB$ and $\mC$ are related by $\mathfrak{st}$ modular transformation, see Fig.~\ref{3tori} in the main text. 

The presented method relies on writing a non-trivial resolution of the identity element of the modular group in terms of its generators. The one that is going to be useful for us reads $(\mathfrak{st})^3 = \bbI$. The cyclic construction requires computing three sets of ground states on tori related by the $\mathfrak{st}$ transformation. It follows \cite{zhang2015general} that the matrices $K,L,M$ in (\ref{eq:KLM}) are expressed by the topological $S$ and $T$ matrices in the following way
\begin{eqnarray}
K &=& (D^\mA)^\dagger ST (D^\mB P^\mB)  \ , \label{Kmat}\\
L &=& (D^\mB P^\mB)^\dagger ST (D^\mC P^\mC) \ , \label{Lmat}\\
M &=& (D^\mC P^\mC)^\dagger ST (D^\mA) \ . \label{Mmat}
\end{eqnarray}
In the above equations, the presence of the diagonal matrices $D^\mA$, $D^\mB$ and $D^\mC$ reflects the fact that every ground state $\vert\Psi_i\rangle$ is accompanied by a random phase. Let us denote those phases by $e^{i\alpha_i}$, $e^{i\beta_i}$ and $e^{i\gamma_i}$ for the three sets of ground states 
\begin{equation}
D^\mA_{ii} = e^{i\alpha_i}, \quad D^\mB_{ii} = e^{i\beta_i}, \quad D^\mC_{ii} = e^{i\gamma_i} \ .
\end{equation}

It is not assumed here that $\ket{\Psi^i_\mA}$, $\ket{\Psi_\mB^i}$ and $\ket{\Psi_\mC^i}$ have the same type of anyon propagating inside the torus (with an exception for the $i=1$ that is assumed to have an identity flux). This lack of knowledge is taken into account by permutation matrices $P^\mB$ and $P^\mC$ in (\ref{Kmat})---(\ref{Mmat}) 
\begin{equation}
P^\mB_{ij} = \delta_{i,p(j)}, \quad P^\mC_{ij} = \delta_{i,q(j)} \ .
\end{equation}
The presented method enables finding permutations $p$ and $q$ (though, only $p$ is required in order to establish $S$ and $T$).

Matrix $P^\mB$ permutes the basis $\{\ket{\Psi_\mB^i} \}$ and $P^\mC$ permutes $\{\ket{\Psi_\mC^i} \}$ in such a way that ground states $\ket{\Psi_\mA^i}$, $\ket{\Psi_\mB^{p(i)}}$ and $\ket{\Psi_\mC^{q(i)}}$ have the same type of anyon flux inside the torus. In our construction, the first element of each basis is required to have an identity flux. This means that $p(1)=q(1)=1$.

Taking the most general form of $S$ and $T$ matrices in (\ref{Kmat})---(\ref{Mmat}), we obtain
\begin{eqnarray}
K_{ij} &=& 	e^{-i\frac{2\pi}{24}c} e^{-i(\alpha_i - \beta_{p(j)})}
			\theta_{p(j)} S_{i,p(j)} \ , \\
L_{ij} &=&	e^{-i\frac{2\pi}{24}c} e^{-i(\beta_{p(i)} - \gamma_{q(j)})}
			\theta_{q(j)} S_{p(i),q(j)} \ , \\
M_{ij} &=&	e^{-i\frac{2\pi}{24}c} e^{-i(\gamma_{q(i)} - \alpha_j)}
			\theta_j S_{q(i),j} \ ,
\end{eqnarray}
where $T_{kk} = e^{-i \frac{2\pi}{24} c} \theta_k$, $c$ is the topological central charge of the anyon model and $\theta_k$ is the twist of an anyon type $k$.

The unknown phases $\alpha,\beta,\gamma$ can be cancelled out by taking the following combination of the matrix elements of $K,L,M$
\begin{eqnarray} \label{U}
U_{ijk}	&\equiv& K_{ij} L_{jk} M_{ki} \\
		&=&		e^{-i\frac{2\pi}{8}c} 
				\theta_i \theta_{p(j)} \theta_{q(k)}
				S_{i,p(j)} S_{p(j),q(k)} S_{q(k),i} \ . \nonumber
\end{eqnarray}

$S$ and $T$ matrices can now be extracted from the elements of $U$  in a following way. First, the central charge $c$ is calculated (modulus 8) from $U_{111}$
\begin{equation} \label{c}
U_{111} = e^{-i\frac{2\pi}{8}c} (S_{11})^3\ ,
\end{equation}
where we have used the fact that $S_{11} > 0$.

Next, twists $\theta_i$ are obtained by examining the elements $U_{i11}$
\begin{equation} \label{theta}
e^{i\frac{2\pi}{8}c} U_{i11} = \theta_i S_{11} (S_{1i})^2 \ .
\end{equation}

$\theta_i$ can be computed from the above equation by recalling that $S_{1i} > 0$.

The information extracted from (\ref{c},\ref{theta}) completely determines the $T$-matrix.

Similarly, permuted entries of the $T$-matrix, $\theta_{p(i)}$ and $\theta_{q(i)}$, are calculated from $U_{1i1}$ and $U_{11i}$
\begin{eqnarray}
e^{i\frac{2\pi}{8}c} U_{1i1} &=& \theta_{p(i)} S_{11} (S_{1,p(i)})^2 \ , \\
e^{i\frac{2\pi}{8}c} U_{11i} &=& \theta_{q(i)} S_{11} (S_{1,q(i)})^2 \ .
\end{eqnarray}

In general, some anyons may have the same twist. Because of that, knowing $\theta_i$, $\theta_{p(i)}$ and $\theta_{q(i)}$ does not always allow to establish permutations $p$ and $q$. This problem can be solved by more careful analysis of the entries of~$U$ that is described below.

To shorten the notation, let us denote
\begin{eqnarray}
V_{ijk}	&\equiv&		e^{i\frac{2\pi}{8}c} \theta_i^* \theta^*_{p(j)} \theta^*_{q(k)} 
					U_{ijk} \nonumber \\
		&=&			S_{i,p(j)} S_{p(j),q(k)} S_{q(k),i} \ .
\end{eqnarray}

The $S$-matrix is extracted from $V$ in the following sequence of steps. First, $S_{11}$ is obtained from $V_{111}$
\begin{equation}
V_{111} = (S_{11})^3 \ .
\end{equation}
Then, the first row of $S$ is calculated from $V_{i11}$
\begin{equation}
V_{i11} (S_{11})^{-1} = (S_{1i})^2 \ .
\end{equation}

Similarly, we get $S_{1,p(i)}$ and $S_{1,q(i)}$ by examining $V_{1i1}$ and $V_{11i}$
\begin{eqnarray}
V_{1i1} (S_{11})^{-1} &=& (S_{1,p(i)})^2 \ , \\
V_{11i} (S_{11})^{-1} &=& (S_{1,q(i)})^2 \ .
\end{eqnarray}

Next, we calculate $S_{i,p(j)}$ from $V_{ij1}$
\begin{equation} \label{Sp}
V_{ij1} (S_{1,p(j)} S_{1,i})^{-1} = S_{i,p(j)} \ .
\end{equation}

After that step, the $S$-matrix is obtained up to the permutation $p$ of its columns. In order to read $p$, we calculate $S_{i,q(j)}$ and $S_{p(i),q(j)}$ from $V_{i1j}$ and $V_{1ij}$ respectively
\begin{eqnarray}
V_{i1j} (S_{1,i} S_{1,q(j)})^{-1} 	&=&	S_{i,q(j)} \ , \\
V_{1ij} (S_{1,p(i)} S_{1,q(j)})^{-1}	&=&	S_{p(i),q(j)} \ .
\end{eqnarray}

The permutation $p$ is now obtained by comparing two matrices $S_{i,q(j)}$ and $S_{p(i),q(j)}$ because they differ by a permutation of rows only. This can always be done since the $S$ matrix is unitary and thus does not have two rows that are the same.

Having calculated the permutation $p$, we read the $S$-matrix from $S_{i,p(j)}$ which was computed in (\ref{Sp}).

So far, the computation of $S$ and $T$ is general and does not require iPEPS description of the ground states. It can be applied to 2D iDMRG simulations in the absence of rotational symmetry and hence it generalizes the method described in \cite{cincio2013characterizing}.

\begin{figure}
\includegraphics[scale=0.6]{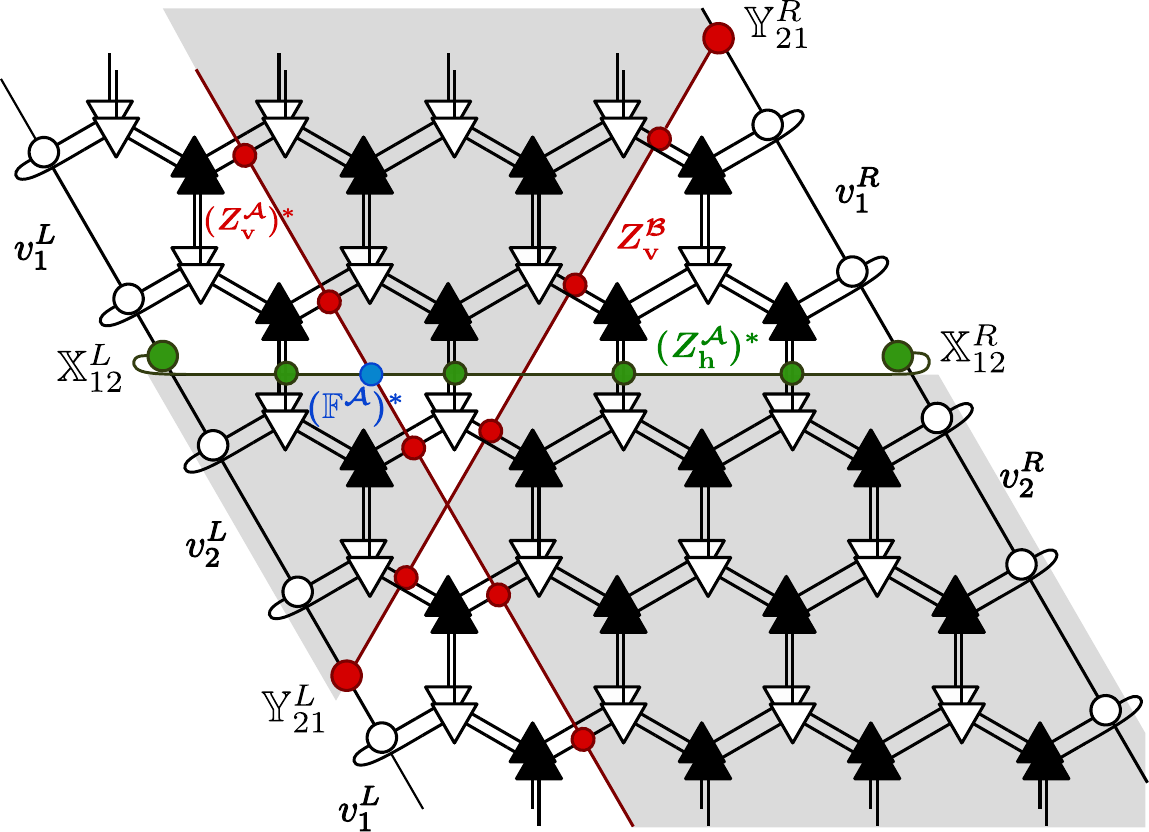}
\caption{ Computation of the last two overlaps in (\ref{nzov}). Impurities (red and green lines) lay only in one layer (either bra or ket). Therefore, impurity fixed points $\bbX_{12}^L$ and $\bbY_{21}^L$ join segments of the boundary iMPS made of $v_1^L$ (white background) and $v_2^L$ (grey background). Superscripts $L$ and $R$ denote left and right eigenvectors of a corresponding transfer matrix, respectively. $\bbY$ is an additional, variationally optimized tensor that was used to find eigenvectors of horizontal impurity transfer matrix, compare Fig.~\ref{fig:impurityTM} in the main text. Furthermore, locations of $\bbY_{21}^L$ along the boundary iMPS is shifted by one unit cell after each application of the impurity TM. 
}
\label{ITM_general}
\end{figure}

Let us now assume that the states $\ket{\Psi^i_\alpha}$ are given by combinations of PEPS with or without insertion of $Z$ generators, as outlined in the main text. Recall that we have:
\begin{align}
	\ket{\Psi^\bbI_\alpha}     &\sim \ket{\Psi^\alpha} + \ket{\Psi_\rmv^\alpha}, &
	\ket{\Psi^e_\alpha}        &\sim \ket{\Psi^\alpha} - \ket{\Psi_\rmv^\alpha}, \\
	\ket{\Psi^m_\alpha}        &\sim \ket{\Psi_\rmh^\alpha} + \ket{\Psi_\rmhv^\alpha}, &
	\ket{\Psi^\epsilon_\alpha} &\sim \ket{\Psi_\rmh^\alpha} - \ket{\Psi_\rmhv^\alpha}
\end{align}
on all tori $\alpha=\mA,\mB,\mC$.
The only non-zero overlaps are those with the same number of $Z$ generators in bra and ket layers of PEPS, both horizontally and vertically. Note that $Z^\mB_\rmv$ extends in the direction which is a linear combination of $w_1^\mA$ and $w_2^\mA$, which means that it winds around the torus $\mA$ once vertically and once horizontally. The same situation occurs for $Z^\mA_\rmh$ when computing overlap with a state on torus $\mB$. 

In what follows, we only analyze overlaps between states on tori $\mA$ and $\mB$. Computation of $L_{ij}$ and $M_{ij}$ in Eq.~(\ref{eq:KLM}) is analogous. The only non-zero contributions to overlaps $\langle \Psi^i_\mA | \Psi^j_\mB \rangle$ are
\begin{equation} \label{nzov}
\langle \Psi^\mA       | \Psi^\mB       \rangle, \
\langle \Psi^\mA_\rmv  | \Psi^\mB_\rmh  \rangle, \
\langle \Psi^\mA_\rmh  | \Psi^\mB_\rmhv \rangle, \ 
\langle \Psi^\mA_\rmhv | \Psi^\mB_\rmv  \rangle \ .
\end{equation}

Calculating the first two overlaps of (\ref{nzov}) is trivial as we already have eigenvectors of iPEPS TM and we can simply insert $Z_\rmv^\mA$ to the bra layer and $Z_\rmh^\mB$ to the ket layer. Because they extend in the same direction, the tensor network that describes the overlap $\langle \Psi^\mA_\rmv | \Psi^\mB_\rmh \rangle$ simplifies:
\begin{align*}
\langle \Psi^\mA_\rmv | \Psi^\mB_\rmh\rangle 
&= \mathrm{Tr}\left(\left(\sum_i \vert v^R_i)(v_i^L\vert\right)^{N_h} (Z_\rmv^\mA)^* Z_\rmh^\mB  \right) \\
&= \sum_{i=1,2}(v_i^L\vert (Z_\rmv^\mA)^* Z_\rmh^\mB  \vert v_i^R) \\
&= \sum_{i=1,2} \ingr{1.3}{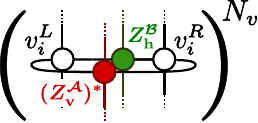} \ , \numberthis \label{eq:ov_ex}
\end{align*}
where $|v_i^R)$ and $(v_i^L|$ denotes right and left eigenvector of iPEPS TM, respectively.

The last two overlaps in (\ref{nzov}) require a generalization of the impurity in TM, as shown in Fig.~\ref{ITM_general}. The impurity lies in one layer only (either bra or ket), so the insertion of an impurity fixed point $\bbX_{ik}, \bbY_{ik}$ changes the boundary iMPS from $v_i$ to $v_k$, where $i\neq k$, hence dividing the whole lattice into two subregions. They are distinguished from each other by different background color in Fig.~\ref{ITM_general}. Furthermore, the location of $\bbY_{ik}$ is shifted by one unit cell after each application of impurity TM to the boundary iMPS. For instance, tensor network resulting from overlap $\langle\Psi^\mA_\rmhv |\Psi^\mB_\rmv\rangle$ can be reduced to

\begin{equation}
\langle\Psi^\mA_\rmhv | \Psi^\mB_\rmv\rangle = \sum_{i\neq k}
\ingr{2.8}{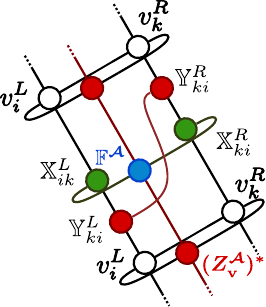} \quad .
\end{equation}

\setcounter{figure}{0} 
\section{String-net models} \label{app:sn}

\begin{figure*}[b]
\begin{center}
\includegraphics[width=0.95\textwidth]{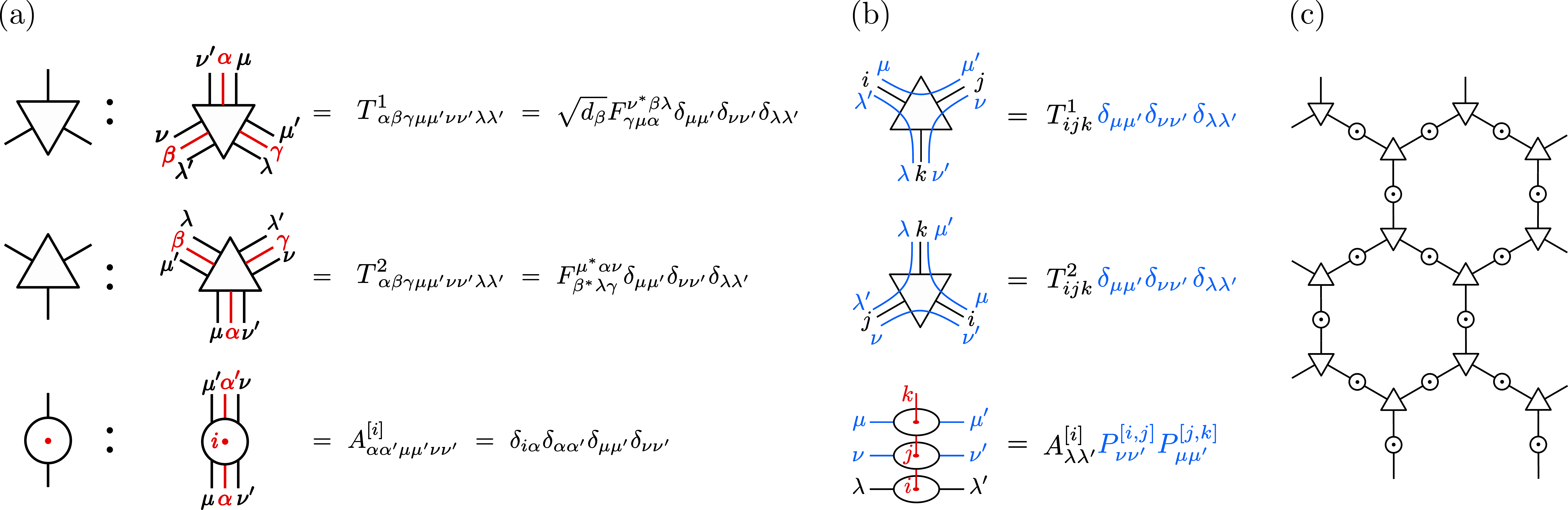}
\caption{
(a) Fixed point PEPS tensors on a honeycomb lattice constructed using $F$-symbols. The physical index is denoted by a red dot on a circular $A$ tensor.  $d_\alpha$ is the quantum dimension of quasiparticle $\alpha$.
(b) Ferromagnetic perturbation applied to the fixed point tensors as a PEPO layer (blue).
(c) The final state is obtained by contracting the virtual indices on a honeycomb lattice.
}
\label{basic_tensors}
\end{center}
\end{figure*}

PEPS describing string-net models consists of the tensors shown in the Fig.~\ref{basic_tensors}, where the physical index is denoted by red dot on a circular $A$ tensor. The symbol $d_\alpha$ is a quantum dimension of a quasiparticle $\alpha$. The total quantum dimension is $D = \sqrt{\sum_\alpha d_\alpha^2}$. The circular tensor $A^i$ with physical index can be incorporated into one of the tensors $T$ hence giving rise to the tensor network with just two types of tensors $\mathbb{A},\mathbb{B}$, defined in Fig. \ref{fig:TM}(A), which we use for clarity of discussion. This construction gives fixed point tensors with zero correlation length. In order to drive the system away from fixed-point, we add a perturbation layer, similar to local filtering considered in \cite{per1,per2,per3}, but instead of single-site string tension we consider a two-site ferromagnetic interaction with the operator $\mathrm{e}^{\lambda \sum_{\langle i,j\rangle}\sigma^x_i\sigma^x_j}$ in order to preserve degeneracy of iPEPS. This also leads to continuous phase transition with divergent correlation length and a ferromagnetic phase at large $\lambda$. The perturbation results in a PEPO layer on top of the fixed point tensors. It can be rewritten as
\begin{eqnarray}
\mathrm{e}^{-\lambda H} &\equiv& 
\prod_{\langle i,j \rangle} \mathrm{e}^{\lambda\sigma^x_i\sigma^x_j} \propto
\prod_{\langle i,j \rangle} \left( \mathbb{I}+\sigma^x_i\sigma^x_j\tanh\lambda \right) \nonumber \\
&=&\prod_{\langle i,j \rangle} \sum_{d_{ij}=0,1}(\mathcal{P}_i\mathcal{P}_j)^{d_{ij}},
\end{eqnarray}
where the operator $\mathcal{P}_i = \sigma^x_i \sqrt{\tanh\lambda}$ and $d_{ij}$ is a value of the bond index connecting nearest neighbor spins $i$ and $j$. It follows that the single operator acting on site $i$ is given by $P_i=\left(\sqrt{\tanh(\lambda)}\sigma^x_i\right)^{d_\mu}$, where $d_\mu$ is a sum of values of blue bond indices connecting the tensor with the perturbation layer as it is shown in Fig.~\ref{basic_tensors}. This construction results in sparse tensors with an unnecessary large bond dimension which can be compressed by a proper svd decomposition.

\setcounter{figure}{0} 
\section{iPEPS energies in Kitaev model} \label{app:var}

\begin{table}[t!]
\begin{tabular}{|c|c|c|c|c|}
\hline
$J_x=J_y$ & $-E_{\rm exact}$ & $-E_{D=4}$ & $-E_{D=5}$ & $-E_{D=6}$ \\ 
\hline
$0.20$ & $1.02015$ & $1.02015$ & $--$  & $--$ \\
$0.30$ & $1.04582$ & $1.04579$ & $--$  & $--$ \\
$0.40$ & $1.08282$ & $1.08260$ & $1.08262$  & $1.08263$ \\
$0.41$ & $1.08719$ & $1.08691$ & $1.08692$  & $1.08694$ \\
$0.42$ & $1.09170$ & $1.09135$ & $1.09141$  & $1.09141$ \\
$0.43$ & $1.09634$ & $1.09593$ & $1.09594$  & $1.09595$ \\
$0.44$ & $1.10112$ & $1.10064$ & $1.10072$  & $1.10072$ \\
$0.45$ & $1.10604$ & $1.10552$ & $1.10561$  & $1.10562$ \\
$0.46$ & $1.11110$ & $1.11061$ & $1.11066$  & $1.11068$ \\
$0.47$ & $1.11632$ & $1.11552$ & $1.11566$  & $1.11569$ \\
$0.48$ & $1.12169$ & $1.12079$ & $1.12097$  & $1.12099$ \\
\hline
\end{tabular}
\caption{ Variational energies in the Kitaev model. }
\label{tab:var}
\end{table}

The ground state of the Kitaev model was obtained with the variational iPEPS algorithm \cite{varCorboz}. We set $J_z=1$ and varied $J_x=J_y$ from $0.40$ to $0.48$. 
The critical point is located at $J_x=J_y=0.5$.
Table \ref{tab:var} lists variational energies for bond dimensions $D=4,5,6$.
The CTMRG bond dimension $\chi=D^2$ was sufficient to reach convergence in $\chi$ for these data.

The full update (FU) scheme \cite{Cirac_iPEPS_08,fu}, that is much easier to implement and more efficient to execute, failed to provide topological ground states. In order to avoid trapping in local minima of energy early in the imaginary time evolution towards the ground state, one has to begin the FU evolution with a relatively large imaginary time step. For a large step, the nearest-neighbor Suzuki-Trotter gates fail to notice the topological constraints that are stabilized by the effective toric code Hamiltonian \cite{kitaev2006anyons} because it is a tiny fourth-order perturbative Hamiltonian acting on whole 6-site plaquettes.


\begin{thebibliography}{65}
\bibitem [{not()}]{note_pure}%
  \BibitemOpen
  \href@noop {} {}\bibinfo {note} {The notions of pure MPSs and pure MPOs are
  introduced in Appendix~\ref{app:tme}.}\BibitemShut {Stop}%
\end{thebibliography}
\end{document}